\documentclass[amsmath,latexsym]{article}
\usepackage{latexsym}
\begin{document}
\begin{center}
{   EXPANDING  $AdS_5$ BRANES:  TIME  DEPENDENT  EIGENVALUE
PROBLEM AND  PRODUCTION  OF  PARTICLES}

\bigskip
\bigskip

I. Brevik\footnote{Email: iver.h.brevik@ntnu.no}, M. Wold Lund,
and G. Ru{\o}

\bigskip
\bigskip

Department of Energy and Process Engineering, Norwegian University
of Science and Technology, N-7491 Trondheim, Norway
\vspace{.5cm}

\begin{abstract}
An analysis is first given of the situation where a scalar field
is contained between two fixed, spatially flat, branes. The usual
fine-tuning (RS) condition is relaxed, and the branes are allowed
to possess a positive effective cosmological constant $\lambda$.
We first analyze the eigenvalue problem for the Kaluza-Klein
masses when the metric is time dependent, and consider in detail
the case when $\lambda $ is small. Thereafter we consider, in the
case of one single brane, the opposite limit in which $\lambda$ is
large, acting in a brief period of time $T$, and present a "sudden
approximation" calculation of the energy produced on the brane by
the rapidly expanding de Sitter space during this period.
\end{abstract}

\vspace{1cm}

June 2004

\end{center}
\newpage

\section{INTRODUCTION}

We begin by considering two static three-branes, situated at fixed
positions $y=0$ and $y=R$ in the transverse $y$ direction. The
branes are embedded in an $AdS_5$ space, and are subject to
fine-tuning. This is the classic scenario of Randall and Sundrum
\cite{randall99}. There are by now several papers on brane
cosmology in general \cite{binetruy00}. If $\Lambda \;(<0)$ is the
five-dimensional cosmological constant in the bulk, the Einstein
equations are
\begin{equation}
R_{AB}-\frac{1}{2}g_{AB}R+g_{AB}\Lambda =\kappa^2 T_{AB},\label{1}
\end{equation}
where $x^A=(\;t,x^1,x^2,x^3,y)$, and $\kappa^2=8\pi G_5$ is the
gravitational coupling. On the first, $y=0$ brane (the Planck
brane) the tensile stress $\tau_0$ is assumed positive. This is
physically tantamount to assuming that the brane contains an ideal
fluid whose equation of state is $p_0=-\rho_0$, i.e., a "vacuum
fluid", but being without any mechanical stress \cite{brevik03}.
Thus $\tau_0$ means physically the same as  $-p_0$.

We introduce the effective four-dimensional cosmological constant
on the two branes,
\begin{equation}
\lambda_0=\frac{1}{6}\Lambda+\frac{1}{36}\kappa^4\tau_0^2, \quad
\lambda_R=\frac{1}{6}\Lambda +\frac{1}{36}\kappa^4\tau_R^2.
\label{2}
\end{equation}
If $\lambda_0 >0$ then $\lambda_R>0$ automatically
\cite{brevik03}. Thus the system is two $dS_4$ branes embedded in
an $AdS_5$ bulk.

We assume that the space is spatially flat, $k=0$. The metric will
be written as \cite{brevik02}
\begin{equation}
ds^2=A^2(y)(-dt^2+e^{2\sqrt{\lambda_0}\,t}\,\delta_{ik}dx^idx^k)+dy^2,
\label{3}
\end{equation}
with $\mu=\sqrt{-\Lambda/6}$, $H_0=\sqrt{\lambda_0}$ being the
Hubble constant on the first brane, and
\begin{equation}
A(y)=\frac{\sqrt{\lambda_0}}{\mu}\sinh [\mu (y_H-|y|)]. \label{4}
\end{equation}
Here $y_H\; (>0)$ is the horizon, determined by the relation
$\sinh (\mu y_H)=\mu/\sqrt{\lambda_0}$. We assume $R<y_H$, so that
the horizon at which $g_{tt}=0$ does not occur in between the
branes. Note that the metric (\ref{3}) reduces to the RS metric
\cite{randall99} in the limit when $\lambda_0=0$.

Consider now a massive scalar field $\Phi$ in the bulk. Its action
can in view of the $Z_2$ symmetry be written as
\begin{equation}
S=\frac{1}{2}\int d^4x \int_{-R}^{R}
dy\sqrt{-G}\,(g^{AB}\partial_A \Phi \partial_B \Phi -M^2\Phi),
\label{5}
\end{equation}
with $G=\det( g_{AB})$.

In Ref.~\cite{brevik01} we calculated the thermodynamic energy at
finite temperatures, in the restrictive case of fine-tuning
($\lambda=0)$. Then, we could make use of the particle concept for
the field  without any conceptual problems. Our purpose in the
present paper is to discuss certain aspects of the more
complicated case when the four-dimensional cosmological constants
$\lambda_0$ and $\lambda_R$ are positive. In the next section we
review briefly for reference purposes the thermodynamic energy
calculation in the fine-tuned case. In Sect. 3 we solve the
eigenvalue problem for the scalar field, in principle. The
calculation is carried out in full for the limiting case when
$\lambda_0$ is small, $\sqrt{\lambda_0}/\mu \ll 1$. Somewhat
surprisingly, in this limit the formalism does not allow any real
value for the Kaluza-Klein masses $m_n$. Finally, in Sect. 4 we
consider the opposite limit in which $\lambda$ is large, but
active during a brief period of time $T$ only. We here restrict
ourselves to the case of one single brane. This situation means
physically a rapid expansion of the de Sitter space, from an
initial static (fine-tuned) case I to another static case II. The
situation is tractable analytically when one makes use of the
"sudden" approximation in quantum mechanics. The production of
particles is estimated from use of the Bogoliubov transformation,
relating the states I and  II. The method was first made use of in
a cosmological context by Parker \cite{parker69}.

\section{ON THE FINE-TUNED CASE}

We assume the branes to be spatially flat, $k=0$, and also to be
fine-tuned, $\lambda_0=\lambda_R=0$. As mentioned avove, we can
here  make use of the particle picture. We assume a finite
temperature, $T=1/\beta$, and start by noting that the free energy
$F$ for a bosonic scalar field in a three-dimensional volume $V$
is given by \cite{brevik01}
\begin{equation}
\beta F=-\ln Z =V\int \frac{d^3p}{(2\pi)^3}\,\ln \left[2\sinh
(\frac{1}{2}\beta E_p)\right], \label{6}
\end{equation}
where $Z$ is the partition function and $E_p=\sqrt{{\bf p}^2+M^2}$
the particle energy.

The metric is now
\begin{equation}
ds^2= e^{-2\mu |y|}\eta_{\alpha\beta}dx^\alpha dx^\beta +dy^2.
\label{7}
\end{equation}
We allow for a boundary mass term by letting $M \rightarrow
\bar{M}$, where
\begin{equation}
{\bar{M}}^2=M^2+2b\mu [\delta(y)-\delta(y-R)], \label{8}
\end{equation}
$b$ being a constant \cite{gherghetta00}. The field equation
 $ (\Box -M^2)\Phi =0 $ in the bulk takes the form
 \begin{equation}
 \left[ e^{2\mu |y|}\,\eta^{\alpha\beta}\partial_\alpha
 \partial_\beta +e^{4\mu |y|}\,\partial_y(e^{-4\mu
 |y|}\partial_y)-M^2\right]\Phi(x^\alpha,y)=0. \label{9}
 \end{equation}
 The Kaluza-Klein decomposition
 \begin{equation}
 \Phi(x^\alpha,y)=\frac{1}{\sqrt{2R}}\sum_{n=0}^\infty
 \Phi_n(x^\alpha)f_n(y) \label{10}
 \end{equation}
 yields the following equation for the KK masses $m_n$:
 \begin{equation}
 e^{-2\mu |y|}\left[ -f_n''(y)+4\; {\rm sign}(y) \mu
 f_n'(y)+M^2 f_n(y)\right]=m_n^2 f_n(y), \label{11}
 \end{equation}
 whose solution (here given for $0<y<R$) can be written as
 \begin{equation}
 f_n(y)=\frac{e^{2\mu
 y}}{N_n}\left[J_\nu\left(x_ne^{\mu(y-R)}\right)+b_nY_\nu\left(x_ne^{\mu
 (y-R)}\right)\right], \label{12}
 \end{equation}
 with $x_n=m_n/(a\mu),\; a=e^{-\mu R},\; \nu=\sqrt{4+M^2/\mu^2}$. The boundary conditions are found
 by integrating Eqs.~(\ref{11}) across the branes. The field may
 be either even (untwisted) or odd (twisted) under the $Z_2$
 symmetry. We define the altered Bessel functions
 $j_\nu(z)=(2-b)J_\nu(z)+zJ_\nu'(z),\;
 y_\nu(z)=(2-b)Y_\nu(z)+zY_\nu'(z)$, and consider here only the even
 case, $f_n(y)=f_n(-y)$. The KK masses are given as roots of the
 equation
 \begin{equation}
 D(x_n) \equiv j_\nu(x_n)y_\nu(ax_n)-j_\nu(ax_n)y_\nu(x_n)=0,
 \label{13}
 \end{equation}
 and the coefficient $b_n$ in Eq.~(\ref{12}) is
 $b_n=-j_\nu(x_n)/y_\nu(n_n)$.

 In the physically reasonable limit of $m_n \ll \mu$ and $a \ll 1$
 we get
 \begin{equation}
 x_n=(n+\frac{\nu}{2}-\frac{3}{4})\pi, \quad n=1,2,..., \label{14}
 \end{equation}
 the approximation being better the larger the value of $n$. The quantities
 $x_n$ are thus for low $n$ of order unity. At
 finite temperatures the bosonic free KK energy $F^{KK}$ can be
 expressed as a contour integral in the following form:
\begin{equation}
\beta F^{KK}=V\int
\frac{d^3p}{(2\pi)^3}\frac{i}{2\pi}\int_Cdx\,\frac{d}{dx}\ln
\left[2\sinh \left(\frac{1}{2}\beta\sqrt{{\bf
p}^2+a^2\mu^2x^2}\right)\right]\,\ln D(x) \label{15}
\end{equation}
for the even  modes, the contour $C$ encompassing the zero points
for $D(x)$ \cite{brevik01}.

This concludes our brief review of the $k=0,\, \lambda=0$ theory,
and we now proceed to consider the non-static case for which
$\lambda$ is different from zero.

\section{TWO BRANES, WHEN $\lambda >0$}

We shall still keep the spatial curvature $k$ equal to zero, but
assume now that $\lambda_0>0$ which, as noted above, implies that
also $\lambda_R>0$. The metric is given by Eqs.~(\ref{3}) and
(\ref{4}). To begin with, we do not restrict $\lambda_0$ to be
small.

The presence of curved branes (meaning here the time varying scale
factor $a=e^{H_0 t}$), complicates the calculation of the vacuum
energy in two ways. First, it results in a non-trivial spectrum of
the KK excitations on the brane. Secondly, for each given
excitation the vacuum energy becomes more complicated because of
the curvature. Actually, the calculation of the determinant of the
relevant differential operator is complicated to such an extent
that its evaluation becomes rather intractable. For this reason,
mostly conformal fields have been investigated so far. Some papers
are listed in Ref.~\cite{nojiri00}; they are dealing with
conformal and partly also with massive fields. Below, we we will
focus attention on one specific issue, namely how to find the
energy eigenvalues of the massive scalar field.

The $y$ equation giving the KK masses now takes the form
\begin{equation}
-f_n''(y)+4\,{\rm sign}(y)\mu \coth [\mu (y-y_H)]f_n'(y)+M^2f_n(y)
=\frac{m_n^2\mu^2}{\lambda \sinh^2 [\mu (y-y_H)]}f_n(y),\label{16}
\end{equation}
for which the solutions are seen to be either even or odd. We
consider here the solution for $y>0$. Introducing the variable $z$
by $z=\cosh \mu (y_H-y)$, we can write the solution as
\begin{equation}
f_n(z)=\frac{1}{N_n(z^2-1)^{3/4}}\left( P_\alpha^\beta
(z)+c_nQ_\alpha^\beta(z)\right), \label{17}
\end{equation}
where $P_\alpha^\beta$ and $Q_\alpha^\beta$ are associated
Legendre polynomials, $\alpha=\sqrt{4+M^2/\mu^2}-1/2,\;
\beta=\sqrt{9/4-m_n^2/\lambda}$, and $N_n$ is a normalization
constant. Note that when $y$ increases from 0 to $R$, $z$
decreases from $z_0$ to $z_R$, where
$z_0=\sqrt{1+\mu^2/\lambda_0},\; z_R=\cosh [\mu (y_H-R)]$. The
Legendre polynomials are usually taken to converge for $|z|<1$,
but can be analytically continued to $|z|>1$, which is the region
of interest here. (To make the functions single valued, a cut is
made along $-\infty <z<1$.)

For {\it odd} solutions we have as boundary conditions
\begin{equation}
f_n\big|_{z_0,z_R}=0, \label{18}
\end{equation}
so that the coefficients $c_n$ in Eq.~(\ref{17}) are determined as
\begin{equation}
c_n=-P_\alpha^\beta
(z_0)/Q_\alpha^\beta(z_0)=-P_\alpha^\beta(z_R)/Q_\alpha^\beta(z_R),
\label{19}
\end{equation}
and the KK masses are given by
\begin{equation}
P_\alpha^\beta(z_0)Q_\alpha^\beta(z_R)-P_\alpha^\beta(z_R)Q_\alpha^\beta(z_0)=0.
\label{20}
\end{equation}
Note that the KK masses enter through the upper index in the
associated Legendre functions.

For {\it even} solutions the boundary conditions are likewise
found by integrating Eq.~(\ref{16}) across the branes. Let us
introduce the notation
\[
p_\alpha^\beta(z)=\frac{1}{z^2-1}\left(\frac{3}{2}z+b\right)P_\alpha^\beta(z)+\left(
P_\alpha^\beta(z)\right)', \]
\begin{equation}
q_\alpha^\beta(z)=\frac{1}{z^2-1}\left(\frac{3}{2}z+b\right)Q_\alpha^\beta(z)+\left(
Q_\alpha^\beta(z)\right)'. \label{21}
\end{equation}
Then, a brief calculation shows that the boundary conditions yield
for the coefficients $c_n$
\begin{equation}
c_n=-p_\alpha^\beta(z_0)/q_\alpha^\beta(z_0)=-p_\alpha^\beta(z_R)/q_\alpha^\beta(z_R),
\label{22}
\end{equation}
and the corresponding KK masses are given by
\begin{equation}
p_\alpha^\beta(z_0)q_\alpha^\beta(z_R)-p_\alpha^\beta(z_R)q_\alpha^\beta(z_0)=0.
\label{23}
\end{equation}
We outline how the KK masses $m_n$ can be calculated, when we
start from the natural assumption that
  $\{\mu, \lambda_0, R\}$ are
known input parameters. The horizon is first determined from the
equation $\sinh(\mu y_H)=\mu/\sqrt{\lambda_0}$, and $z_0$ and
$z_R$ are found from the expressions above. If moreover the scalar
mass $M$ is known, the value of $\alpha$ follows, and the KK
masses finally follow by solving Eqs.~(\ref{20}) and (\ref{23}) in
the odd and even cases, respectively.

Due to the appearance of $m_n$ in the upper index $\beta$, the
formalism becomes however quite complicated. We shall not here
carry on the analysis further, except from pointing out the
following and perhaps surprising  behavior of the formalism in the
limiting case of a small-$\lambda_0$, narrow-gap situation. Let us
first assume that $\lambda_0/\mu \ll 1$. Then,
$z_0=\sqrt{1+\mu^2/\lambda_0}=\cosh (\mu y_H) \gg 1$. Moreover, if
we take the gap width $R$ to be small compared with the horizon,
$R\ll y_H$, then we can also assume that $ z_R=\cosh [\mu (y_H-R)]
\gg 1$. It becomes thus natural to make use of the following
approximate expressions for the analytically continued Legendre
functions, valid when $|z| \gg1$:
\begin{equation}
P_\alpha^\beta(z)=\left\{ \frac{2^\alpha \Gamma
(\alpha+\frac{1}{2})}{\sqrt{\pi}\Gamma(\alpha-\beta+1)}z^\alpha
+\frac{\Gamma(-\alpha-\frac{1}{2})}{2^{\alpha+1}\sqrt{\pi}\Gamma(-\alpha-\beta)}z^{-\alpha-1}\right\}
 \left(1+O\left(\frac{1}{z^2}\right) \right). \label{24}
\end{equation}
This expression holds when $2\alpha \neq \pm1,\pm3,...$.
Similarly,
\begin{equation}
Q_\alpha^\beta(z)=\sqrt{\pi}\frac{e^{\beta \pi i}}{2^{\alpha+1}}
\frac{\Gamma(\alpha+\beta+1)}{\Gamma(\alpha+\frac{3}{2})}z^{-\alpha-1}\left(
1+O\left(\frac{1}{z^2}\right)\right), \label{25}
\end{equation}
which holds when $2\alpha \neq  -3,-5,...$ \cite{gradshteyn80}. In
the case of odd solutions, insertion of Eqs.~(\ref{24}) and
(\ref{25}) in Eq.~(\ref{20}) leads to the condition
$1/\Gamma(\alpha-\beta+1)=0$, which means $\alpha-\beta+1=-n$ with
$n=0,1,2...$. In other words
\begin{equation}
\sqrt{\frac{9}{4}-\frac{m_n^2}{\lambda}}=n+\frac{1}{2}+\sqrt{4+\frac{M^2}{\mu^2}}.
\label{26}
\end{equation}
This is a condition, however, that cannot be satisfied for any $n
\geq 0$. The case of even solutions leads to the same conclusion.
Thus in the limit of small $\lambda_0$ and narrow gap $R$ there is
no physical solution for $m_n$.

\section{ENERGY PRODUCTION CALCULATED VIA THE BOGOLIUBOV TRANSFORMATION}

Instead of dealing with the complicated case of a time dependent
metric, it is sometimes possible to proceed in another way which
is both mathematically simple and physically instructive. The
method applies in cases where the change of the system takes place
abruptly (corresponding to the "sudden approximation" in quantum
mechanics). The method implies use of the Bogoliubov
transformation relating two vacua, designated in the following by
I (in) and II (out). As mentioned in Sect. 1, in a cosmological
context the method was introduced by Parker \cite{parker87}, and
it has been made use of later, for instance in connection with the
 formation of cosmic strings \cite{parker87,brevik95}.

We assume now that there is only one single brane, situated at
$y=0$,  As before, we take $k=0,\lambda_0>0$,
 and  we start from the state I in which the metric is static
and the particle concept for the field thus directly applicable.
 This state corresponds  to
times $t<0$. The line element on the brane at these times is thus
$ds_I^2=-dt^2+\delta_{ik}dx^idx^k$. During the time period $0<t<T$
the de Sitter metric on the brane is rapidly expanding, and the
line element is $ds^2=-dt^2+e^{2H_0 t}\delta_{ik}dx^idx^k$. For
$t>T$ (state II) the metric is again assumed static, so that
$ds_{II}^2=-dt^2+\alpha^2\delta_{ik}dx^idx^k$, with
$\alpha=e^{H_0T}$. For  state II the particle concept is thus
again applicable. The vacua corresponding to states I and II are
denoted by $|0,I\rangle$ and $|0,II \rangle$.

The field equation on the brane reads, in both static cases,
\begin{equation}
\ddot{\Phi}_n-\partial^i\partial_i\Phi_n+m_n^2\Phi_n=0, \label{27}
\end{equation}
and in state I the basic modes can be written
\begin{equation}
u_{{\bf k}I}=\frac{(2\pi)^{-3/2}}{\sqrt{2\omega_I}}\exp(i\,{\bf
k}_I\cdot {\bf x}-i\omega_I t), \label{28}
\end{equation}
with ${\bf k}_I=(k_1,k_2,k_3)_I$. The dispersion relation is
$\omega_I^2={\bf k}_I^2+m_n^2$. In state II the basic modes
$u_{{\bf k}II}$ are given by the same expression, only with the
replacements $\omega_I\rightarrow \omega_{II},\; {\bf
k}_I\rightarrow {\bf k}_{II}$. Here ${\bf k}_{II}=(k_1, k_2,
k_3)_{II}$ and $\omega_{II}^2=\alpha^{-2}{\bf k}_{II}^2+m_n^2$
(note that
$k_{II}^ik_{iII}=\alpha^{-2}k_{iII}k_{iII}=\alpha^{-2}{\bf
k}_{II}^2$). In both cases the mode functions satisfy the
normalization condition
\begin{equation}
(u_{\bf k},u_{\bf k'}) \equiv -i\int u_{\bf k}(x)
\stackrel{\leftrightarrow}{\partial_0}u_{\bf k'}(x)\,d^3
x=\delta({\bf k}-{\bf k'}), \label{29}
\end{equation}
the other products vanishing, as usual. We consider the same
positions in space before and after the expansion, i.e.,
$x_I^\mu=x_{II}^\mu \equiv x^\mu$. Correspondingly,  we identify
the covariant components of the wave vectors, $k_i^I=k_i^{II}
\equiv k_i$. The dispersion relations then become
\begin{equation}
\omega_I^2={\bf k}^2+m_n^2,\quad \omega_{II}^2=\alpha^{-2}{\bf
k}^2+m_n^2. \label{30}
\end{equation}
We expand $\Phi_n$ in either I or II modes,
\begin{equation}
\Phi_n(x)=\int \frac{d^3 k}{(2\pi)^3}\left( a_{\bf k}u_{\bf
k}(x)+a_{\bf k}^\dagger u_{\bf k}^* (x)\right), \label{31}
\end{equation}
and relate the two modes via a Bogoliubov transformation
\cite{birrell82}
\begin{equation}
u_{{\bf k}II}=\int \frac{d^3k'}{(2\pi )^3}\left( \alpha_{\bf k
k'}\,u_{{\bf k'}I}+\beta_{\bf k k'}\,u_{{\bf k'}I}^*\right),
\label{32}
\end{equation}
implying for the operators
\begin{equation}
 a_{{\bf k}I}=\int
\frac{d^3 k'}{(2\pi)^3}\left(\alpha_{\bf k' k}\,a_{{\bf
k'}II}+\beta_{\bf k' k}^* \, a_{{\bf k'}II}^\dagger \right).
\label{33}
\end{equation}
The number of particles $N(k)$ in $\bf k$ space in state II,
produced by the rapid expansion, is
\begin{equation}
N(k)=\langle 0,II| a_{{\bf k}I}^\dagger a_{{\bf k}I} |0,II\rangle
=\int \frac{d^3k'}{(2\pi)^3}|\beta_{\bf k' k}|^2, \label{34}
\end{equation}
where we calculate
\begin{equation}
\beta_{\bf k' k}=-(u_{{\bf k'}II}, u_{{\bf
k}I}^*)=\frac{1}{2}\left(\sqrt{\frac{\omega_I}{\omega_{II}}}
-\sqrt{\frac{\omega_{II}}{\omega_I}}\right)\,\delta(\bf k + \bf k'
). \label{35}
\end{equation}
We insert this into Eq.~(\ref{34}), and make use of the effective
substitutions
\begin{equation}
\int \frac{d^3k'}{(2\pi)^3}\left[ (2\pi)^3\delta({\bf k} + {\bf
k'})\right]^2 \rightarrow (2\pi)^3\delta({\bf k} +{\bf
k'})\big|_{{\bf k'} \rightarrow {\bf -k}} \rightarrow V,
\label{36}
\end{equation}
where $V$ is the volume, to get
\begin{equation}
N(k)=\frac{V}{4(2\pi)^3}\left(
\frac{\omega_I}{\omega_{II}}+\frac{\omega_{II}}{\omega_I}-2\right).
\label{37}
\end{equation}
Multiplying with $\omega_{II}$ to get the energy of each mode, we
find by integrating over all $\bf k$ the following simple
expression for the produced energy:
\begin{equation}
E=\pi V \int_0^K \omega_{II}\left(
\frac{\omega_I}{\omega_{II}}+\frac{\omega_{II}}{\omega_I}-2\right)
k^2 dk. \label{38}
\end{equation}
Here $K$ is introduced as an upper limit, to prevent the UV
divergence arising from the idealized sudden approximation.

We shall consider only one limit of this expression. As noted in
Sect. 2, $x_n =(m_n/\mu)e^{\mu R} \sim 1$ usually, implying that
$m_n /\mu \ll 1$. Let us simply assume here that $m_n=0$, so that
$\omega_I/\omega_{II} =\alpha$ according to Eq.~(\ref{30}). Then,
Eq.~(\ref{38}) yields
\begin{equation}
E=\frac{1}{4}\pi V K^4\left(
1+\frac{1}{\alpha^2}-\frac{2}{\alpha}\right). \label{39}
\end{equation}
To make a rough estimate of the energy produced we may, following
Parker \cite{parker87}, take into account that the process
actually takes some finite time $\Delta t$. The consequence is
that the spectrum falls rapidly for frequencies much larger than
$1/\Delta t$. Let us simply assume that $K$ is of the same order
of magnitude as $1/\Delta t$. Then, Eq.~(\ref{39}) yields in
dimensional units
\begin{equation}
E \sim \frac{\pi \hbar V}{4c^3(\Delta
t)^4}\left(1+\frac{1}{\alpha^2}-\frac{2}{\alpha}\right).
\label{40}
\end{equation}
If we take $\alpha$ to be of order unity, and set $\Delta t \sim t
\sim 10^{-33}$ s (i.e., inflationary times), then this expression
yields $E/V \sim 10^{73}$ ergs/${\rm cm^3}$. For the sake of
comparison, we note that this is about 13 orders of magnitude less
than the energy density associated with the formation of a cosmic
string \cite{parker87}.

\end{document}